\documentstyle[twocolumn,aps,psfig,floats]{revtex}           

\begin{document}                          	     

\def\be{\begin{equation}}
\def\ee{\end{equation}}
\def\vnab{{\vec\nabla}}
\draft                                    	     

\twocolumn[\hsize\textwidth\columnwidth\hsize\csname 
@twocolumnfalse\endcsname

\title{Multiscale Finite-Difference-Diffusion-Monte-Carlo Method for Simulating Dendritic Solidification}

\author{Mathis Plapp and Alain Karma}

\address{
Physics Department and Center for Interdisciplinary Research
on Complex Systems, \\
Northeastern University, Boston, Massachusetts 02115 
}

\date{January 11, 2000}

\maketitle

\begin{abstract}
We present a novel hybrid computational method
to simulate accurately dendritic solidification in the
low undercooling limit where the dendrite tip radius 
is one or more orders of magnitude smaller than the 
characteristic spatial scale of variation of the 
surrounding thermal or solutal diffusion field. 
The first key feature of this method is an efficient multiscale 
diffusion Monte-Carlo (DMC) algorithm which allows off-lattice random 
walkers to take longer and concomitantly rarer steps with increasing
distance away from the solid-liquid interface. 
As a result, the computational cost of evolving the 
large scale diffusion field becomes insignificant when
compared to that of calculating the interface evolution. 
The second key feature is that random walks are only permitted
outside of a thin liquid layer surrounding the interface. Inside
this layer and in the solid, the diffusion equation is solved 
using a standard finite-difference algorithm that is interfaced 
with the DMC algorithm using the local conservation law for 
the diffusing quantity.
Here we combine this algorithm with a previously developed 
phase-field formulation of the interface dynamics and demonstrate
that it can accurately simulate three-dimensional dendritic growth
in a previously unreachable range of low undercoolings
that is of direct experimental relevance.
\end{abstract}
]

\section{Introduction}
\nobreak

Diffusion-limited pattern formation, which leads to the
spontaneous emergence of complex branched structures, 
occurs in numerous contexts. A few examples include 
dendritic solidification \cite{Kurz}, electrochemical deposition
\cite{eldep} and corrosion, 
and the growth of bacterial colonies \cite{BenJacob}.
Two distinct length scales are typically involved in this
class of problems: one that characterizes 
the pattern itself, such as the thickness of a 
branch, and one that characterizes the diffusion 
field associated with the transport of heat or matter. 
In many cases, these two scales are vastly different.
For example, in solidification, the decay length of
the thermal or solutal field ahead of a growing
dendrite (in a pure or alloy melt) can be one to three orders
of magnitude larger than the tip radius of one of its primary 
branches. Non-trivial pattern formation dynamics  
can be expected to occur on all intermediate scales.
This poses a serious challenge for numerical
simulations since a precise integration of
the equations of motion on the pattern scale requires a good resolution 
of the interfacial region, and such a resolution is completely inefficient
(i.e. much too fine) to treat the large scale 
diffusion field. Therefore, in order to retain this precision
on the small scale and, at the same time, simulate the pattern evolution
on sufficiently large length and time scales, it is 
necessary to use some form of multiscale algorithm. 

Multi-grid and finite element methods with non-uniform meshing
represent one possible solution for this type of problems.
Their application, however, in the context of growth
simulations faces the additional difficulty of a
moving interface, which implies that the structure 
of the simulation grid has to be dynamically adapted.
For the classic problem of dendritic 
crystal growth, several multi-grid \cite{Braun97}
or adaptive meshing algorithms \cite{Schmidt96} have been
proposed in recent years. The most precise to date is the method
of Provatas {\it et al.} which uses the phase-field model
on a regular grid on the scale of the dendrite,
whereas the diffusion field is integrated on an adaptive
mesh using finite element techniques \cite{Provatas99}. 
While this method appears to be promising, it has yet 
to be implemented in three dimensions where the difficulty
of adaptive meshing becomes significantly enhanced.

We present in this paper an alternative solution to
solve this computational challenge and we illustrate its
application in the context of
the dendritic crystallization of a pure substance
from its undercooled melt, even though this algorithm can 
be applied to any diffusion-limited growth problem 
for which an explicit solver of the interface 
dynamics is available. The idea is to use a hybrid
approach. The interface dynamics is treated using
deterministic equations of motion, in particular 
those of the phase-field model for the dendritic
growth problem considered here.
On the other hand, the large-scale diffusion field
is represented by an ensemble of off-lattice random 
walkers and is evolved using a Diffusion Monte Carlo 
(DMC) algorithm. The two solutions are connected at 
some distance from the moving interface.
The key point for rendering our method efficient is that
we use random walkers which dynamically adapt the average 
length of their random steps. Far from the interface,
the walkers can make large jumps and hence be updated
only rarely without affecting the quality of the solution
near the growing interface. In some sense, our
method can be seen as an ``adaptive grid algorithm
without grid''. The DMC algorithm and the connection
between deterministic and stochastic parts are rather
simple and straightforward to implement in both two 
and three dimensions, both on single-processor and
parallel architectures. We demonstrate in this paper
that our method is precise, robust, and reliable,
and hence constitutes a powerful alternative to
state of the art adaptive meshing techniques.
Technically, the algorithm bears 
many similarities to quantum Monte Carlo methods 
\cite{Koonin}. It is therefore remarkable 
that the gap between mesoscopic and macroscopic length scales 
can be bridged using a method borrowed from 
microscopic physics in an interfacial pattern 
formation context, which was not {\it a priori} obvious to 
us at the start of this investigation.

Our algorithm builds on ideas of earlier 
random walk algorithms for simulating pattern 
formation during viscous fingering \cite{Kadanoff85,Liang86} 
and solidification \cite{Vicsek84,Nittmann86,Karma87,Saito89},
but introduces two essential new features. Firstly,
random walks with variable step size have been used
previously in simulations of large-scale diffusion-limited
aggregation \cite{Meakin83}, but only one walker at a
time was simulated, and hence the time variable did not
explicitly appear in the treatment of the walkers.
In the present diffusive case, the memory of the past 
history, stored in the diffusion field, is essential to 
the problem. Our DMC algorithm works with a whole ensemble
of walkers in ``physical'' time and hence constitutes a 
true multiscale solver for the full diffusion problem.
Secondly, the algorithms mentioned above use a lattice
both to evolve the walkers and to represent the position
of the interface by the bonds between occupied (solid)
and empty (liquid) sites. Walkers are created or absorbed 
directly at this interface. The discretization of space and 
the stochastic creation and absorption of walkers make it 
difficult to control accurately the interfacial anisotropy 
and the noise that both play a crucial role in
dendritic evolution \cite{Nittmann86,Brener96}.
Consequently, the algorithms aimed at describing dendritic 
growth \cite{Vicsek84,Nittmann86,Saito89},
while correctly reproducing all the qualitative features
of the growth process, are unable to yield quantitative
results that can be tested against experiments. We solve 
both problems by creating and absorbing walkers not at the
solid-liquid interface, but at a ``conversion boundary''
at some fixed distance from the interface. This means
that the stochastic representation of the diffusion
and the motion of the interface can be treated separately,
which allows us to evolve the interface accurately
by the phase-field method using a finite difference
representation of controlled precision.
At the same time, the stochastic noise created by the
DMC algorithm is rapidly damped by the deterministic diffusion
in the ``buffer layer'' between the conversion boundary
and the solid-liquid interface, and hence the amplitude of the 
fluctuations {\em at the solid-liquid interface} can be reduced 
to a prescribed level without much cost in computation
time by increasing the thickness of the buffer layer.
This is an important issue for simulations of dendritic growth, 
because the amplification of microscopic fluctuations
of the interface is believed to be the main cause for the
formation of secondary dendrite branches \cite{Langer87},
and it is well known that numerical noise can lead to the
formation of spurious sidebranches in simulations. 
Consequently, we have to demonstrate that the walker 
noise of our algorithm can be reduced to a level that does 
not affect the pattern evolution.

Another benefit of the buffer layer is that it makes the 
algorithm very versatile. Away from the interface, only the 
standard diffusion equation has to be solved. Therefore, the 
DMC part of the algorithm and the conversion process between
deterministic and stochastic solutions are completely
independent from the method used for simulating the
interface dynamics, and can easily be carried over to
other free boundary problems.

The purpose of the present paper is to describe the 
algorithm in detail and to demonstrate its reliability
and precision by benchmark simulations. Some results
concerning three-dimensional crystal growth at low
undercoolings have already been presented elsewhere
\cite{Plapp99,Tip99}, and hence we will focus here
on the computational aspects of the problem. 
Section II contains a brief introduction to dendritic 
solidification and the basic equations of motion,
and describes the phase-field method. In section III, 
the DMC algorithm and its interfacing with the 
phase-field equations are described in detail.
In section IV, we present results of benchmark simulations, 
assess the efficiency of the code and the magnitude
of numerical noise, and present
simulations of three-dimensional dendritic growth.
Section V contains a conclusion and the outline of future work.

\section{Dendritic growth and the phase-field method}
\nobreak

When a crystal grows from an undercooled melt, it develops
into an intricate branched structure, called a dendrite.
This phenomenon has been of central importance to the
understanding of spontaneous pattern formation during
phase transformations and the emergence of branched
structures \cite{Langer80,Kessler88,Brener91}.
In addition, it is of considerable practical 
interest, because dendrites form during the
solidification of many commercially important alloys
and influence the mechanical properties of the 
finished material.

We will focus on the dendritic solidification 
of a pure substance from its homogeneously undercooled 
melt, starting from a single supercritical nucleus 
\cite{Huang81,Rubinstein91,Muschol92,Glicksman94,Bisang95}. 
This situation is well described by the
symmetric model of solidification, which assumes
that the diffusivity and thermophysical quantities 
such as the specific heat and the density are equal for 
the solid and the liquid phases.
During the growth of the crystal, the latent
heat of melting is released, and in the absence
of convection, the growth becomes limited by the
diffusion of heat away from the growing dendrite.
The state of the system at any time is described
by the temperature field $T(\vec x,t)$ and the
shape $\Gamma(t)$ of the boundary between solid 
and liquid. It is customary to define a dimensionless 
temperature field
\be
u(\vec x,t) = {T(\vec x,t)-T_m\over L/c_p},
\ee
where $L$ and $c_p$ are the latent heat of melting
and the specific heat, respectively, and $T_m$ is the
melting temperature. In terms of this field, the 
equations of motion of the symmetric model are
\be
\partial_t u = D{\vnab}^2 u,
\label{diffuu}
\ee
\be
v_n = D \hat n\cdot\left({\vec\nabla u |}_S - 
                         {\vec\nabla u |}_L\right),
\label{stefanu}
\ee
\be
u_\Gamma = - d_0 \,\sum_{i=1}^{d-1}\,
\left[a(\hat n)+\frac{\partial^2a(\hat n)}
{\partial \theta_i^2}\right]{1\over R_i}-\beta(\hat n) v_n,
\label{githou}
\ee
where $D$ is the thermal diffusivity, $v_n$ is the
normal velocity of the interface, and $\hat n$ 
is the unit normal vector of the surface $\Gamma$ 
pointing towards the liquid. The diffusion equation,
Eq. (\ref{diffuu}) is valid everywhere (in the liquid
and in the solid) except on the surface $\Gamma$.
The Stefan condition, Eq. (\ref{stefanu}), valid on
$\Gamma(t)$, expresses the conservation of enthalpy
at the moving phase boundary. Here, $\vnab u|_S$
and $\vnab u|_L$ denote the limits of the
temperature gradient when $\Gamma$ is approached from the
solid and the liquid side, respectively, and the
equation states that the local heat flux at the
interface must be equal to the latent heat generated
or consumed during the phase transformation; $v_n$
is positive if the solid grows (i.e. freezing).
The dimensionless temperature at the interface $u_\Gamma$ is
given by the generalized Gibbs-Thomson condition Eq. (\ref{githou}).
The first term on the right hand side (RHS) is the anisotropic 
form of the local equilibrium condition (Gibbs-Thomson
condition) which relates the temperature to the curvature
of the interface and the anisotropic surface tension
$\gamma(\hat n)=\gamma_0 a(\hat n)$. For a crystal with
cubic symmetry in three dimensions, the anisotropy
function $a(\hat n)$ is usually written as
\be
a(\hat n) = (1-3\epsilon_4)\,
\left[1+\frac{4\epsilon_4}
{1-3\epsilon_4}\left(n_x^4+n_y^4+n_z^4\right)\right],
\label{ani}
\ee
where $\epsilon_4$ is the anisotropy parameter.
Note that in two dimensions ($d=2$), this expression 
reduces to
\be
a(\theta) =  1 + \epsilon_4 \cos(4\theta),
\label{ani2d}
\ee
where $\theta$ is the angle between the normal and one 
of the axes of symmetry. On the RHS of Eq. (\ref{githou}),
\be
d_0 = {\gamma_0 T_m c_p\over L^2}
\ee
is the capillary length, $d$ is the spatial dimension,
$\theta_i$ are the angles between the normal $\hat n$ 
and the two local principal directions on $\Gamma$, 
and $R_i$ are the principal radii of curvature. 
Finally, the second term on the RHS of Eq. (\ref{githou})
describes the shift of the interface temperature 
due to molecular attachment kinetics, and $\beta(\hat n)$
is the orientation-dependent linear kinetic coefficient.
Kinetic effects are believed to be small for the range of
solidification speeds of interest here. We will
therefore focus on the case where the interface 
kinetics vanish ($\beta(\hat n) \equiv 0$), which 
corresponds to local equilibrium at the interface.
In this case, the physical length and time scales are 
set by the capillary length and the diffusivity,
and the control parameters of the problem are
the anisotropy $\epsilon_4$ and the dimensionless 
undercooling
\be
\Delta = {T_m - T_0\over L/c_p},
\ee
where $T_0$ is the initial temperature, $T(\vec x,0)=T_0$,
which provides the thermodynamic driving force for 
solidification. We assume that the dendrite grows into an 
infinite volume of liquid, and hence $u(\vec x,t)\to -\Delta$
as $|\vec x|\to \infty \, \forall \,t$. Typical experimental
values for $\Delta$ range from $0.001$ to $0.1$. 
The length scales involved in the problem 
are (i) the capillary length $d_0$,
(ii) a typical scale of the pattern such as the radius
of curvature at a tip $\rho$, and (iii) the length scale of the
diffusion field $l_D$. To fix the ideas, let us consider
the measurements of Rubinstein and Glicksman on pivalic
acid (PVA) \cite{Rubinstein91}. For a dimensionless undercooling 
of $\Delta=0.075$, $\rho = 8.5 \,\mu m$, and the speed of the
tips is $v=390 \,\mu m/s$, which gives a diffusion length
$l_D = 2D/v = 0.38 \,mm$, whereas $d_0 = 3.8 \,nm$. The
multiscale character of this situation is obvious: 
$l_D$ and $d_0$ differ by five
orders of magnitude, and $l_D$ is forty times larger 
than $\rho$. These ratios become even larger for lower
undercoolings.

The above equations constitute a notoriously difficult
free boundary problem. To simplify the task, theoretical and 
numerical efforts first concentrated on the treatment of a
single needle crystal growing at constant velocity.
This situation can be treated by boundary integral
methods \cite{Kessler88}, which are exact in two
dimensions (2-d) but have remained approximate in three
dimensions (3-d). More recently, time-dependent methods 
have been developed to describe the full growth dynamics
\cite{Sethian92,Ihle94,Karma98}. Of those,
the phase-field method \cite{pf} seems presently
the most compact and precise approach. We use
a recent efficient formulation of this method,
which has been benchmarked against boundary
integral calculations \cite{Karma98}.
An ``order parameter'', or phase-field 
$\psi(\vec x,t)$ is introduced, which is an indicator 
field distinguishing the solid ($\psi=1$) and
the liquid ($\psi=-1$) phase.
The two-phase system is described by a free energy
functional of Ginzburg-Landau type,
\be
{\cal F}=\int dV\,[\,W^2({\hat n})|\vnab\psi|^2+f(\psi,u)],
\label{freeen}
\ee
where $W({\hat n})$ is the orientation-dependent interface
thickness, i.e. the spatial scale on which the phase-field
varies smoothly between its equilibrium values $\psi=\pm 1$,
and $f(\psi,u)$ is the free energy density. The
equations of motion are
\be
\tau(\hat n)\partial_t \psi = -{\delta {\cal F} \over \delta \psi(\vec x,t)},
\label{pfpsi}
\ee
where $\delta{\cal F}/\delta\psi$ denotes the functional derivative, and
\be
\partial_t u = D\,\vnab^2 u+\frac{1}{2} \partial_t \psi.
\label{pfu}
\ee
The phase-field relaxes to its local minimum free energy 
configuration, which depends on the local temperature
field, with an orientation-dependent relaxation time 
$\tau(\hat n)$. The diffusion equation contains a source
term to account for the latent heat released or consumed
during the phase transformation.
For a suitable choice of the functions $f(\psi,u)$, 
$W(\hat n)$ and $\tau(\hat n)$, these equations 
reduce precisely to the free boundary
problem given by Eqs. (\ref{diffuu}) to (\ref{githou})
in the limit where the interface thickness is small
compared to the radii of curvature \cite{Karma98}. A brief
description of the model used for our simulations
and its relation to the macroscopic free boundary problem
is given in the appendix. The key point is that the
phase-field equations of motion are partial
differential equations which can be integrated on
a regular grid on the scale of the dendrite, without
knowing explicitly where the solid-liquid interface
is located. The phase field rapidly decays to its
equilibrium values $\psi=\pm 1$ away from the interface.
Therefore, well within the bulk phases, Eq. (\ref{pfpsi}) 
becomes trivial and Eq. (\ref{pfu}) reduces to the 
ordinary diffusion equation.

\section{Diffusion Monte Carlo algorithm}
\nobreak
\subsection{Outline}
\nobreak
\noindent
Our goal is to combine the precision of the 
phase-field method and the efficiency
of a DMC treatment for the diffusion field. 
\begin{figure}
\centerline{
\psfig{file=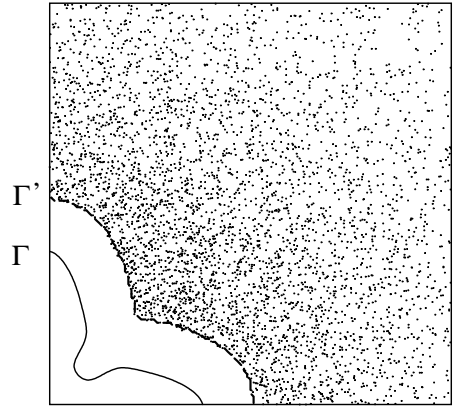,width=.4\textwidth}}
\caption{Simulation of two-dimensional dendritic 
growth for a dimensionless undercooling $\Delta=0.1$
and a surface tension anisotropy  
$\epsilon_4=0.025$. The solid line is the solid-liquid 
interface $\Gamma$, the dashed line is the conversion boundary $\Gamma'$
between the inner (deterministic) and outer (stochastic) domains, 
and the dots show the positions of random walkers (only one walker
out of 50 is shown for clarity).}
\label{figillu}
\end{figure}
This is achieved by dividing the 
simulation domain into an ``inner'' and 
an ``outer'' region as shown in Fig. \ref{figillu}. 
In the inner region, consisting of the growing structure and 
a thin ``buffer layer'' of liquid, we integrate the 
phase-field equations described above. In the outer 
region, the diffusion field is represented by an ensemble 
of random walkers. Walkers are created and absorbed at the 
boundary between inner and outer domains at a rate which
is proportional to the local diffusion flux. 
The value of the diffusion field in the outer 
domain is related to the local density of walkers, 
and the boundary conditions for the integration in 
the inner region are obtained by averaging this density 
over coarse-grained boxes close to the boundary.
We will now describe in detail the DMC algorithm
for the evolution of the random walkers and the
connection of the two solutions.

Let us start by recalling some well-known facts about random
walkers. Consider first a single point particle performing
a Brownian motion in continuous space and time. The conditional
probability $P({\vec x}',t'|\vec x,t)$ of finding 
the particle at position ${\vec x}'$ at time $t'$, 
given that it started from position $\vec x$
at time $t$, is identical to the diffusion kernel,
\be
P({\vec x}',t'|\vec x,t) = 
   {1\over \left[4\pi D(t'-t)\right]^{d/2}}
     \exp{\left[-{|{\vec x}'-{\vec x}|}^2\over 4D(t'-t)\right]},
\label{diffker}
\ee
where $D$ is the diffusion coefficient and $d$ is the
spatial dimension. This kernel satisfies the well-known
convolution relation
\begin{eqnarray}
P({\vec x}'',t''|\vec x,t) &=& 
  \int\,P({\vec x}'',t''|{\vec x}',t')P({\vec x}',t'|\vec x,t)\,d{\vec x}'
   \nonumber\\
  && \quad\forall\, t<t'<t''.
\label{convo}
\end{eqnarray}
Therefore, a realization of a random walk, i.e. the position
of a walker as a function of time, represented by a time-dependent
vector of real numbers $\vec x(t)$, can be obtained on a 
computer by successive steps. The position of the walker 
is updated following the scheme
\be
\vec x(t+\tau) = \vec x(t) + \ell \vec \xi,
\label{scheme}
\ee
where the components of the random vector $\vec\xi$ are
independent Gaussian random variables of unit variance.
The time increment $\tau$ (not to be confused with 
the phase-field relaxation time $\tau(\hat n)$ 
defined in the preceding section) and the
step size $\ell$ must satisfy the relation
\be
{\ell^2\over\tau} = 2 D.
\label{taucond}
\ee
Since time is continuous and Eq. (\ref{convo}) is not
restricted to $t''-t'=t'-t$, successive 
steps may have different time increments (and concomitantly
use different step lengths) if Eq. (\ref{taucond})
is satisfied for each update.

The basic idea of Diffusion Monte Carlo simulations is to
sample many realizations of diffusion paths. The density
of random walkers then satisfies a stochastic differential
equation which converges to the deterministic diffusion
equation in the limit of an infinite number of walkers.
A density of walkers can be defined by a suitable 
coarse-graining procedure on a scale $L_{cg}$, i.e.
by dividing space into cells of volume $L_{cg}^d$ and
counting the number of walkers within each cell. 
If the coarse-graining length is chosen larger than
the average step length $\ell$, this density evolves
smoothly on the scale of $L_{cg}$ over times of order
$L_{cg}^2/D$, the time for one walker to diffuse
through a coarse cell.

From the above considerations, it is clear that the
characteristic length and time scales that can be
resolved by a stochastic DMC algorithm are set by
the step size $\ell$ and the time increment $\tau$,
respectively. The key point is that for the present
application a high spatial and temporal resolution
is needed {\em only close to the interface}, whereas
far from the dendrite, the coarse-graining length
and hence the step size can become much larger than
the fine features of the growing crystal. In practice,
we choose the step size to be approximately proportional
to the distance $d_{cb}$ of the walker from the conversion
boundary between the inner (deterministic) and outer 
(stochastic) regions, i.e.
\be
\ell \approx c\, d_{cb}
\label{ldist}
\ee
with a constant $c\ll 1$. According to Eq. (\ref{taucond}), 
the time increment between updates grows as the square 
of the step size, and hence the walkers
far from the dendrite have to be updated only rarely.
We use dynamical lists to efficiently handle the updating
process, as will be described in more detail in Sec. \ref{secwalk}.
For low undercoolings, where the scale of the diffusion field
is much larger than the dendrite itself and most of
the walkers need only be updated sporadically,
we obtain enormous savings of computational 
time over a straightforward integration of
the diffusion equation.

Let us now discuss how the inner and outer regions
are interfaced. Two essential goals have to be accomplished.
Firstly, we have to supply a boundary condition at the
conversion boundary for the integration of the deterministic
equations in the inner region, and secondly we need to
create and absorb walkers at a rate which is proportional
to the local heat flux across this boundary. 

\begin{figure}
\centerline{
\psfig{file=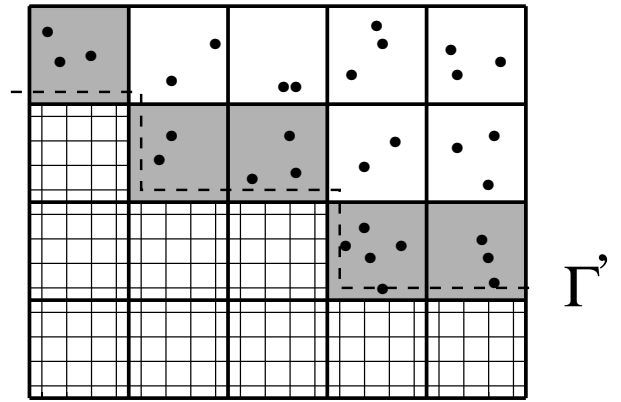,width=.45\textwidth}}
\medskip
\caption{Sketch of a small part of the conversion boundary in
two dimensions for $n=4$. Each cell of the coarse grid (thick
lines) contains $16$ points of the fine grid (thin lines). 
The fine grid is shown only in the inner region for clarity.
The shaded cells are conversion cells, and walkers are 
represented by black dots. The boundary $\Gamma'$ between
inner and outer regions is indicated by a dashed line.}
\label{figgrids}
\end{figure}
The phase-field equations are
integrated in the inner region on a regular cubic grid,
henceforth called ``fine grid'', with spacing $\Delta x$.
Each node on this grid contains the local values of the
phase field $\psi$ and the temperature field $u$.
We superimpose on this grid another, coarser grid,
of mesh size $L_{cg}=n\Delta x$, such that the links 
of the coarse grid intersect the links of the fine grid
as shown in Fig. \ref{figgrids}. The first purpose of
this grid is to define the geometries of the two
simulation regions and of the conversion boundary.
We describe the ``state'' of each coarse cell by an
integer status variable $S_{\alpha\beta\gamma}^t$. 
Here and in the following, greek indices ($\alpha$,
$\beta$, $\gamma$) label the {\em cells} of the coarse grid 
along the $x$-, $y$-, and $z$-directions, whereas
latin indices ($i$, $j$, $k$) label the {\em nodes}
of the fine grid. All cells which contain at
least one node of the fine grid where $\psi>0$ are
assigned the status ``solid'' ($S=-2$). 
All cells with a center-to-center distance to the 
nearest solid cell smaller than a prescribed length 
$L_b$ are ``buffer cells'' ($S=-1$),
whereas all other cells belong to the outer region.
Cells of the outer region which have at least one nearest
neighbor with buffer status are called {\em conversion
cells} ($S=0$) and play the central role in interfacing the
two solutions. The dividing surface $\Gamma'$ between
inner and outer regions is the union of all the links 
(or plaquettes in three dimensions) of the coarse grid 
which separate conversion from buffer cells 
(see Fig. \ref{figgrids}). Evidently, as the crystal
grows, the geometry of the two regions changes, which
means that the status variables must be periodically updated.
Details on this procedure are given in Sec. \ref{secupdate}.

We always choose $L_b$ sufficiently large to ensure that
the phase field is already close to its liquid equilibrium 
value, $\psi \approx -1$, at the conversion boundary. 
Hence we can set $\psi=-1$ in the entire outer region
and treat only the standard diffusion equation there. 
In the initial state, the entire system is undercooled to
$u=-\Delta$, and no walkers are present. When the
crystal grows, it releases latent heat which diffuses
away from the interface, and hence the inner region 
becomes a heat source for the outer region. This heat 
flux is converted into walkers, each walker representing
a certain discrete amount of heat. We define
in each coarse cell an integer variable 
$m_{\alpha\beta\gamma}^t$ which contains the number
of walkers being within this cell at time $t$. For a specific
heat which is independent of temperature, the density
of walkers is proportional to the difference between
the actual and the initial temperatures, i.e. the
temperature in the outer region is related to the
number of walkers by
\be
u_{\alpha\beta\gamma}^t = 
   -\Delta\left(1-{m_{\alpha\beta\gamma}^t\over M}\right),
\label{coarsetemp}
\ee
where the constant $M$ fixes the number of walkers in a
cell that corresponds to the melting temperature $u=0$.

The inner region is completely delimited by conversion
cells. To fix the boundary condition for the integration 
on the fine grid, it is therefore sufficient to set the 
field $u$ on all nodes of the fine grid in each conversion 
cell to the value specified by Eq. (\ref{coarsetemp}).
The diffusion equation is then timestepped in the inner
region using the standard explicit scheme
\begin{eqnarray}
u_{ijk}^{t+\Delta t} & = & u_{ijk}^t + {D\Delta t\over (\Delta x)^2}
   \nonumber \\
  & & \mbox{}\times
   \left(u_{i+1jk}^t+u_{i-1jk}^t+u_{ij+1k}^t+u_{ij-1k}^t\right. \nonumber \\
  & & \quad\mbox{} +
         \left.u_{ijk+1}^t+u_{ijk-1}^t-6u_{ijk}^t\right).
\label{diffudis}
\end{eqnarray}
Note that we have omitted for simplicity the source terms due
to the phase field, which are zero at the conversion boundary. 
Seen on a discrete level, this equation can be interpreted
as a ``pipe flow'' equation: the local change of $u$ is given
by the sum of the ``flow'' through all the discrete links 
(``pipes''), where, for example, the ``flow'' through a link 
along $x$ during a timestep is given by 
$D\Delta t(u_{i+1jk}-u_{ijk})/(\Delta x)^2$.
For nodes at the boundary of the inner region, some links
cross the conversion boundary $\Gamma'$, which means that
there is exchange of heat with the neighboring conversion
cell. This heat flux is collected by the conversion
cell and stored in a heat reservoir variable 
$H_{\alpha\beta\gamma}^t$. A symbolic manner to describe
the updating of $H_{\alpha\beta\gamma}^t$ is
\be
H_{\alpha\beta\gamma}^{t+\Delta t} = H_{\alpha\beta\gamma}^t
  + {D\Delta t\over (\Delta x)^2} 
  \left(\sum_{\rm bonds} u_{\rm grid}^t-u_{cc}^t\right),
\ee
where the sum runs over all the bonds of the fine grid 
that cross $\Gamma'$, $u_{\rm grid}$ is the temperature 
on a node of the fine grid and $u_{cc}$ is the temperature 
in the conversion cell given by Eq. (ref{coarsetemp}). 
For example, for a conversion cell $(\alpha,\beta,\gamma)$ 
in contact with a buffer cell $(\alpha-1,\beta,\gamma)$, 
we have (we recall that the linear dimension of a coarse
cell is $L_{cg}=n\Delta x$):
\begin{eqnarray}
 & & H_{\alpha\beta\gamma}^{t+\Delta t} = 
       H_{\alpha\beta\gamma}^t + \nonumber \\
 & & \quad\sum_{j=(\beta-1)n+1}^{\beta n}\,\, \sum_{k=(\gamma-1)n+1}^{\gamma n}
     {D\Delta t\over (\Delta x)^2} (u_{i-1jk}^t-u_{ijk}^t) \nonumber \\
 & & \quad {\rm with}\quad i=(\alpha-1)n +1.
\label{resupdate}
\end{eqnarray}
If the stored quantity of heat exceeds a critical value $H_c$
given by
\be
H_c = {n^d \Delta \over M},
\ee
a walker is created at the center of the conversion cell
and $H_c$ is subtracted from $H_{\alpha\beta\gamma}$.
Conversely, if the local heat flux is negative (heat
is locally flowing {\em towards} the dendrite) and
$H_{\alpha\beta\gamma}$ falls below $-H_c$,
a walker is removed from the cell and $H_c$ is added to
the reservoir. This algorithm exactly conserves the
total heat if the contributions of the fine grid, the
reservoir variables and the walkers are added.
In dimensional quantities, each walker
is equivalent to an amount of heat $\Delta Q$ equal to
\be
\Delta Q = {L (n\Delta x)^d \Delta \over M}. 
\ee

The walkers are restricted to the outer region. If a
walker attempts to jump across the conversion boundary,
the move is discarded and the walker stays at its old 
position until the next update. If $c$ in 
Eq. (\ref{ldist}) is small enough, such jumps are 
attempted almost only by walkers close to
the conversion boundary. Accordingly, this procedure is
a convenient way of implementing the re-absorption
of walkers: if a walker stays in a conversion cell,
the heat flux is more likely to be directed towards
the inner region, which increases the chances for
the walker to be absorbed. An alternative method,
namely to deposit all the heat contained in a walker
in the fine grid and remove the walker upon its crossing
of the boundary, would create stronger temperature
fluctuations on the fine grid close to the conversion
boundary.

In summary, the conversion process is handled using three
auxiliary fields on the coarse grid: the status field
$S_{\alpha\beta\gamma}$ which encodes the geometry of 
the buffer layer and the conversion boundary, 
the field $m_{\alpha\beta\gamma}$ that contains the number
of walkers in each cell and is zero in the inner region, 
and the heat reservoir field $H_{\alpha\beta\gamma}$, which
is different from zero only in conversion cells. 
Let us comment on the size of the grids and the resulting
memory usage. The fine grid needs to be large enough 
to accommodate the dendrite and the liquid buffer layer 
during the whole time of the simulation. Especially in 
three dimensions, the restrictions on storage space make 
it necessary to fully use the fine grid.
The coarse grid needs to cover at least the same space
region as the fine grid. As will be detailed below,
for an efficient handling of the walkers close to the 
conversion boundary, it is desirable to always have some
portion of coarse grid in front of the conversion boundary,
and hence the coarse grid should actually cover a 
slightly larger region of space than the fine grid. 
Since the coarse grid has far less nodes than the fine 
grid ($1$ node of coarse grid for $n^d$ nodes of fine grid), 
this does not significantly increase the storage requirement.
In addition, we need an array to store the positions of
the walkers. The latter are represented by ``continuous'' 
positions and need no grid for their evolution. The
walkers can therefore leave the region of space where the 
grids are defined and diffuse arbitrarily far away from the
dendrite, allowing us to simulate growth into an infinite medium.
The most storage-intensive part is the fine grid. In fact,
the limiting factor for most of our three-dimensional
simulations is not so much computation time, but rather the
storage space needed to accommodate large dendrites.

Finally, let us describe how the different parts of the 
algorithm are connected. The program runs through the
following steps:
\begin{enumerate}
\item Setup (or update) the status field $S_{\alpha\beta\gamma}$
on the coarse grid to fix the geometry of the conversion boundary 
\item Calculate the temperature in each conversion cell and set
the boundary condition for the inner region on the fine grid
\item Timestep the phase-field equations on the fine grid and
calculate the heat flux between the inner region and the conversion
cells
\item Update the heat reservoir variables $H_{\alpha\beta\gamma}$
and create or absorb walkers in the conversion cells
\item Advance the walkers
\item Repeat steps 2 through 5. From time to time, extract the
shape of the dendrite and store it for future processing. If
the phase boundary has moved by more than a coarse cell size,
go back to step 1.
\end{enumerate}

In the following subsections, we will give more details on
some features of our implementation, such as the updating 
of the walkers, the updating of the geometry, the choice
of parameters, and parallelization.

\subsection{Updating random walkers}
\label{secwalk}
\nobreak
\noindent
Before going into details, let us briefly point out
similarities and differences between our method and
other DMC algorithms. Such methods are widespread in
Quantum Monte Carlo (QMC) calculations where they are used
to solve the Schroedinger equation in imaginary time \cite{Koonin}.
Each walker represents a configuration in a usually
high-dimensional Hilbert space, and the density of
walkers is proportional to the square amplitude of
the wave function. In contrast, in
our method the walkers evolve in real space, and their
density represents the temperature field.
The most important difference, however, is that in QMC
all walkers are usually updated at the same time, whereas
in our method some walkers are updated much more rarely
than others. Therefore, it would be very inefficient 
to visit every walker in each timestep. Instead, we 
work with dynamical lists.

\begin{figure}
\centerline{
\psfig{file=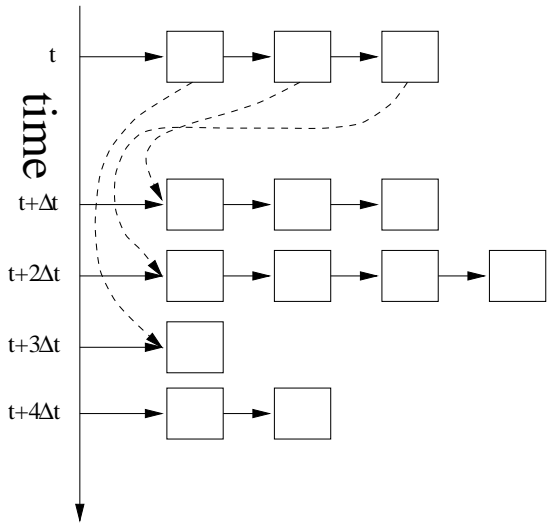,width=.4\textwidth}}
\caption{Sketch illustrating the configuration of the dynamical 
walker lists. Each box stands for a walker, and the full arrows
indicate pointer variables; the ``backbone'' array of pointers
is represented by the downward arrow on the left. At time $t$, 
walkers are updated and prepended to the lists corresponding to
their next update time, as indicated by broken arrows.}
\label{figwalk}
\end{figure}

To simplify the bookkeeping of the different update times,
we enforce that updating takes place only at the discrete
times when the fine grid is updated, i.e. for $t=i\Delta t$,
$i=1,2,\ldots$. Then, we can make a list for every timestep
containing all the walkers that have to be updated at that
moment. However, these lists greatly vary in length and
can therefore not easily be accommodated in standard arrays 
of variables. Therefore, we define a data structure that 
contains the coordinates of one walker plus a pointer 
variable. Within a given list, the pointer associated with one
walker indicates the next element of the list, or contains
an end of list tag if the corresponding walker is the last
one of the list. An array of pointer variables indicates
for each timestep the first element of the corresponding
list. This array is the ``backbone'' of the list structure.
It is easy to add new walkers to a list: the pointer
of the new walker is set to the former first element of
the list, and the pointer of the backbone is set to the
new walker (see Fig. \ref{figwalk}). Lists of arbitrary 
length can be constructed, and every walker is visited 
only when it actually has to be updated.

At a given time $t$, the program works through the
corresponding list of walkers. The treatment of each 
walker starts by looking up the status of the coarse 
grid cell corresponding to its position. If the walker 
is inside a buffer cell because the conversion boundary 
has moved since its last update, it is removed.
This removal does not violate heat conservation because
the heat associated with the walker 
is accounted for in the initialization of the temperature
field inside newly created buffer cells
(see Sec. \ref{secupdate}, Eq. (\ref{initialnewbuff}) below).
If it is inside a conversion cell, and the corresponding 
reservoir variable $H_{\alpha\beta\gamma}^t<-H_c$,
the walker is removed and $H_c$ is added to the 
reservoir. In all other cases, the jump distance
$\ell$ and corresponding time increment are 
determined and a new position is selected according
to Eq. (\ref{scheme}). To apply Eq. (\ref{ldist}) 
for the jump distance $\ell$, we need to determine
the distance of a walker to the conversion boundary.
It would be very inefficient to calculate this distance
for each walker separately, especially when the shape
of the boundary becomes complex. Therefore, we use the
status field $S^t_{\alpha\beta\gamma}$ on the coarse grid in 
the outer region to store an approximate value for this 
distance, which can then be easily looked up by each walker
before a jump. Some more details are given in Sec. \ref{secupdate}.

As mentioned above, we restrict the walker updating to
a discrete set of times. Therefore, the time increment 
$\tau$ in Eq. (\ref{scheme}) has to be an integer
multiple of the time step $\Delta t$, which would not 
be the case if we directly applied Eqs. (\ref{ldist}) 
and (\ref{taucond}). We solve this problem by defining 
a lower cutoff for the jump distances,
\be
\ell_{\rm min} = \sqrt{2Dn_t\Delta t}
\label{lmin}
\ee
where $n_t$ is a fixed integer, and replace the jump 
distances $\ell$ found from Eq. (\ref{ldist}) by
the closest integer multiple of $\ell_{\rm min}$.
We also define a maximum jump length $\ell_{\rm max}$,
mainly to limit the size of the backbone pointer array:
with a maximum jump distance $\ell_{\rm max}$, each
walker is at least updated every $\ell^2_{\rm max}/( 2 D\Delta t)$
timesteps. Consequently, the discrete time modulo this number
can be used to index the pointer variables in the backbone array.

It should be mentioned that in our list structure, it is
difficult to find a walker which is close to a given
position, because all sublists must be searched. This is 
important because the number of walkers in the conversion
cells has to be known for the interfacing with the inner
solution. To avoid time-consuming sweeps through the walker
lists, we update the walker number field $m^t_{\alpha\beta\gamma}$
on the coarse grid whenever a walker jumps.

\subsection{Updating the geometry}
\label{secupdate}
\nobreak
\noindent
We now describe more in detail how the status field on the
coarse grid is setup and adapted to the changing geometry.
When the dendrite grows, the configuration of the buffer layer 
and the conversion boundary has to change in order to maintain a 
constant thickness $L_b$ of the buffer layer. Cells which are 
part of the outer region at the beginning of the simulation 
may become conversion cells, then part of the buffer layer,
and finally part of the dendrite. Under the conditions
we want to simulate, the crystal may locally melt
back, but no large regions of space
will undergo the transition from solid to liquid, and
hence we do not consider the inverse status change 
(from buffer to conversion cell, for example). 
Typically, at low undercoolings a readjustment of the 
geometry becomes necessary only after $1000$ to $10000$
timesteps. Therefore, the efficiency requirements are
not as stringent as in the other parts of the program.

The procedure starts with a sweep through the fine grid. 
Every cell of the coarse grid which contains at least one 
node of the fine grid where $\psi>0$ is assigned 
the status ``solid'' ($S_{\alpha\beta\gamma}^t=-2$).
Next, the solid cells at the boundary of the dendrite
(i.e. each solid cell which has at least one neighboring cell
which is not solid) are used to define the buffer region:
all cells with a center-to-center distance less than $L_b$ 
of a boundary cell which are not solid are assigned the 
status ``buffer'' ($S_{\alpha\beta\gamma}^t=-1$). 
When a conversion cell or a cell of the outer 
region becomes a buffer cell, we need to define the initial
values of the two fields on the fine grid. The phase
field is set to its liquid value, $\psi=-1$. The 
temperature is calculated from the total heat 
contained in the cell, taking into account both the 
walkers and the heat reservoir variables in the conversion 
cells in order to ensure that the total amount of heat 
remains conserved, i.e.
\be
u_{\rm init}={\Delta\over M}\left(m^t_{\alpha\beta\gamma} + 
     H^t_{\alpha\beta\gamma}/H_c\right).
\label{initialnewbuff}
\ee
All nodes of the fine grid within the new buffer cell 
are initially assigned this value. The walkers contained in 
the cell are removed.

All cells of the outer region which are adjacent to the buffer, 
i.e. which have at least one neighbor with buffer status, are
conversion cells ($S_{\alpha\beta\gamma}^t=0$). When a cell 
of the outer region becomes a conversion cell, its heat 
reservoir variable is initialized at zero.

Finally, in the outer region, which is comprised of all
the other cells, the status field is used to store an
approximate value for the distance from the conversion
boundary. A precise determination of this
distance is rather costly in computation time,
because for each cell in the outer region, we 
must calculate the distance to all conversion cells 
and retain the minimum value. A much cheaper, albeit 
approximate method is the following. As mentioned,
in a conversion cell we have $S_{\alpha\beta\gamma}^t=0$.
We assign to all cells adjacent to a conversion cell 
the value $S_{\alpha\beta\gamma}^t=1$. Neighbors
of the latter receive the value $S_{\alpha\beta\gamma}^t=2$,
and we continue this process outward by assigning the value
$S_{\alpha\beta\gamma}^t=i+1$ to all cells adjacent
to a cell with $S_{\alpha\beta\gamma}^t=i$. For a relatively 
simple geometry such as a single growing dendrite, the
status field can be correctly set up on the whole lattice
during a single outward sweep, starting from the center
of the dendrite. The number assigned to a given cell
can be used as a measure for the distance. Note that
the exact relationship of the number to the distance
depends on the direction with respect to the axes of
the coarse grid; our numerical tests below show, however, 
that this anisotropy in the distance function does not
significantly influence the dendrite shapes.

If we follow this procedure, the coarse grid needs to
cover the entire region of space where the jump distance
varies. Even though we introduce a large-scale cutoff
$\ell_{\rm max}$, this would become prohibitive in terms
of memory usage for truly multi-scale problems. Fortunately,
such a sophisticated scheme for the determination of the
distance is mainly needed close to the dendrite (for example,
a walker that enters in the space between two dendrite
arms needs to make small steps). Once a walker has left
the vicinity of the dendrite, this rather complicated
estimate for the distance to the conversion boundary can 
be replaced by a simpler one, for example the distance 
to the closest dendrite tip. In consequence, the coarse
grid needs to cover only a slightly larger region of
space than the fine grid.

Finally, let us comment on the integration of the
phase-field equations in the inner region. We need to
know which part of the fine grid must be timestepped.
This information is encoded in the status field
$S_{\alpha\beta\gamma}^t$ on the coarse grid.
It would, however, be rather inefficient in terms of 
memory access time to integrate the inner region 
``coarse cell by coarse cell''. Instead, integration 
proceeds along the spatial direction
corresponding to successive memory locations, which is
the x-direction in our implementation. During the updating
of the status field, the program determines for each $y$ and $z$
coordinate the range(s) to be integrated along $x$ and keeps
this information in a lookup table. This table is updated
every time the status field changes.

\subsection{Choice of computational parameters}
\nobreak
\noindent
There are number of parameters in our algorithm which
can be adjusted to maximize the computational efficiency.
However, certain restrictions apply. Firstly, there are
various length scales. In order of increasing magnitude,
those are:
\begin{enumerate}
\item the lattice spacing of the fine grid, $\Delta x$,
\item the minimum jump length of the walkers, $\ell_{\rm min}$,
\item the size of a coarse-grained cell, $L_{cg}=n\Delta x$, and
\item the buffer thickness $L_b$.
\end{enumerate}
The minimum jump length should be of the order of the
inner grid spacing to assure a precise interfacing between
inner and outer solutions. On the other hand, a larger
$\ell_{\rm min}$ means less frequent walker updating.
We usually worked with $\ell_{\rm min}\approx 2\Delta x$,
or $n_t\approx 10$ in Eq. (\ref{lmin}). On the other hand,
$\ell_{\rm min}$ has to be smaller than $L_{cg}$ in order
to achieve a well-defined coarse-graining. The 
coarse-graining length, in turn, is limited by geometrical
constraints. The conversion boundary appears ``jagged'' on 
the scale of $L_{cg}$ (see Fig. \ref{figgrids}). In order
to render the effects of this coarse geometry irrelevant
for the interface evolution, the buffer thickness must
be much larger than this scale, $L_b\gg L_{cg}$. We
found that $L_{cg}\approx 0.1\,L_b$ is sufficient to
achieve this goal. In our simulations, we mostly worked
with $n=4$ ($L_{cg}\approx 2\ell_{\rm min}$) and $n=8$
for larger buffer sizes. 

Next, consider the constant of proportionality $c$ between
the walker jump length and the distance to the conversion
boundary, $d_{cb}$. Since the Gaussian random vector $\vec \xi$
in Eq. (\ref{scheme}) has no cutoff, steps of arbitrary length
are possible, and hence even a walker which is far away
can jump directly to the conversion boundary. The number
of such events has to be kept small, because otherwise
the conversion process is influenced by the far field
with its coarse length and time scales. This goal can be
naturally achieved by choosing $c$ small enough. For 
example, for $c=0.1$, only jumps with a length of more
than $10$ standard deviations can reach the conversion
boundary, which represents a negligible fraction. On
the other hand, the increase of $\ell$ with distance
determines the efficiency of the algorithm, and hence
$c$ should be chosen as large as possible. We usually
worked with $c=0.1$, which seems to provide a good 
compromise.

Finally, the parameter $M$ determines the number of
walkers per coarse cell and hence the precision of the
stochastic representation for the temperature field and
the diffusion equation. Considering Eq. (\ref{coarsetemp}),
we see that the temperature at the boundary of the inner
region takes discrete values spaced by $\Delta/M$. In addition,
for a homogeneous distribution of walkers in a system
at $u=0$, the temperature fluctuations are of order
$\Delta/\sqrt{M}$. On the other hand, increasing $M$ means
longer computation time because more random walks have
to be performed. In addition, the total number of walkers $N$
necessary to simulate a dendrite of final volume $V$ is
\be
N = {M V\over (n\Delta x)^d\Delta},
\ee
which means that high values of $M$ become prohibitive,
especially at low undercooling. Fortunately, a good precision
of the solution can be obtained also by increasing $L_b$,
as will be described in Sec. \ref{sectests}. In practice,
we worked with values of $M$ ranging between $25$ and $100$.

\subsection{Boundary conditions and symmetries}
\nobreak
\noindent
For a two-dimensional dendrite seeded at the origin and
with arms growing along the $x$- and $y$-directions, the
simulations can be accelerated by taking advantage of the
cubic symmetry. There are several symmetry axes, and
consequently it is sufficient to integrate the equations
in a part of the plane while imposing reflective boundary
conditions at the proper axes to enforce the symmetry.
These boundary conditions have to be imposed both on the
fine grid and for the walkers. For the symmetry axes at 
$x=0$ and $y=0$, this can be easily achieved by choosing
one of the nodes of the coarse grid to coincide with the
origin. Then, the two symmetry axes coincide with bonds
in the coarse grid. On the fine grid, the nodes
outside the simulation domain but adjacent to the boundary
are set to the values of their mirror images inside the
simulation domain after each timestep. Walkers that attempt 
to cross the boundaries are reflected, i.e. instead of their
``true'' final position outside the simulation domain, its
mirror image with respect to the symmetry axis is chosen. 
Another interpretation of this boundary condition for the 
walkers is to imagine that there exists an ensemble of 
``mirror walkers'' which are the images of the walkers 
inside the simulation domain. When a walker jumps outside
of the simulation domain, its mirror image jumps inside, 
and interchanging the walker and its mirror, we just 
obtain a reflection of the walker at the boundary as above.

\begin{figure}
\centerline{
\psfig{file=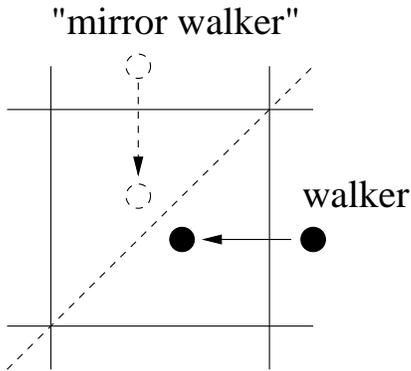,width=.3\textwidth}}
\smallskip
\caption{Sketch illustrating the implementation of reflecting
boundary conditions at the symmetry axis $x=y$. Shown is a
cell of the coarse grid (solid lines) on the diagonal $x=y$
(dashed line). A walker inside the simulation domain ($x>0$, 
$0<y<x$) enters the cell. An accompanying ``mirror walker'' 
(open circles), the image of the walker with respect to the 
symmetry axis, enters the same cell.}
\label{figmirror}
\end{figure}
The latter view is useful when considering the last symmetry
axis, the diagonal $x=y$. While the boundary conditions on
the fine grid and for the walkers can be implemented as before,
the conversion process requires special attention, because
the symmetry axis does not coincide with the boundaries
of a coarse cell. When a walker enters a coarse cell situated
on the diagonal, there is an additional ``mirror walker''
entering {\em the same} coarse cell (see Fig. \ref{figmirror}), 
and hence the number of walkers $m^t_{\alpha\beta\gamma}$ 
has to be increased by two 
(or, equivalently, decreased by two if a walker leaves the cell). 
Similarly, walkers are created and absorbed in
pairs, which means that walker creation in such a cell can
occur only when the heat reservoir exceeds twice the equivalent
of one walker. In addition, when calculating the heat flux
received by conversion cells on the diagonal, both the ``real''
and the ``mirror'' flux has to be taken into account.
It is clear that this procedure induces an
anisotropy in the conversion process; our tests showed,
however, that its effect is undetectable for reasonable
buffer thickness.

In three dimensions, the reduction in computational resources
is even more dramatic. For example, using the symmetry planes
$y=0$, $x=y$, and $x=z$, i.e. integrating only the domain
$x>0$, $0<y<x$, $z>x$, we need only integrate $1/48$ of the
full space, i.e. one eighth of one dendrite arm. The planes
$x=y$ and $x=z$ can be handled as described above, with the
exception of cells on the diagonal $x=y=z$. Such cells
actually have only $1/6$ of their volume within the simulation
domain, and for each walker entering a cell, there are $5$
mirror walkers to be considered.

\subsection{Parallelization}
\nobreak
\noindent
Even though our algorithm is very efficient as will be shown
below, the demands on computation time and RAM storage space
rapidly increase when the undercooling is lowered. Therefore,
we have developed a parallel version of our code for the 
Cray T3E at the National Energy Research Scientific Computing
Center (NERSC), using the shared memory library SHMEM.

We are mainly interested in the development of a single
primary dendrite branch. Hence, an efficient method
of parallelization is to divide the simulation domain in
``slices'' normal to the growth direction, and to distribute
the slices among the processors. In the inner region, the
integration of the partial differential equations makes it
necessary to exchange the boundary values between neighboring
processors after each timestep. This is a standard procedure.
The more delicate points are the handling of the walkers
and the updating of the geometry.

Each processor stores only the parts of the fine grid 
it has to integrate, along with the values of the status
field in the whole simulation domain. The latter is
necessary to correctly handle the walkers. For the
walkers which are far from the dendrite, the average
jump distance may become much larger than the thickness
of a computational slice. But if a walker approaches 
the conversion boundary, the conversion process has to
be handled by the ``local'' processor which contains the
appropriate part of the fine grid. Therefore, the walkers
need to be redistributed after their jumps. We have found
it sufficient to implement ``exchange lists'' between
neighboring processors, i.e. processors which contain
adjacent parts of fine grid. If a walker jumps to a
position outside of the local slice, it is stored
in one of two lists, corresponding to ``upward'' and 
``downward'' motion. After each timestep, these lists
are exchanged between neighboring processors. As most
of the walkers make several small steps before reaching
the conversion boundary, this procedure assures the
correct redistribution of walkers with insignificantly 
few errors, which arise in the rare case that a walker
arrives at the conversion boundary after several large 
jumps.

The only step of the algorithm which needs massive exchange
of data between the processors is the updating of the 
geometry: each processor has to determine locally the
``solid'' part of its computation domain, and this
information has to be exchanged in order to correctly
setup the whole status field on each processor. However,
as mentioned earlier, the geometry is updated only rarely,
and therefore this part of the algorithm does not
represent a significant computational burden. We have
found that the parallel version of our code showed
satisfactory execution time scaling when the number of 
processors is increased.

\section{Numerical tests}
\label{sectests}
\nobreak
\noindent
The accuracy of the standard phase-field method has been
assessed in detail by comparison to boundary integral 
results \cite{Karma98}. Therefore, to test the stochastic
algorithm it is sufficient to check its results against
direct simulations of the standard deterministic 
phase-field equations. The most critical questions
are whether the use of the rather coarse lattice for the 
conversion introduces spurious anisotropy, and what is
the magnitude of the temperature fluctuations generated
by the stochastic treatment of the far field. The main
parameters which control both of these effects are the
thickness of the buffer layer and the number $M$ of walkers
generated per coarse cell. The boundary condition
for the inner region is imposed on a coarse geometry
with a cutoff scale of $n\Delta x$, and the temperature
at the boundary is a stochastic variable which changes
as walkers are created, absorbed, enter, or leave a 
conversion box, and which assumes discrete values spaced 
by $\Delta/M$. When the buffer layer is much larger than 
the size of a coarse cell, $L_b \gg n\Delta x$,
the field is ``smoothed out'' in space and time by the
diffusive dynamics. We expect high spatial and temporal
frequencies to decay rapidly through the buffer layer,
and hence the evolution of the interface to become
smoother as $L_b$ is increased.

\begin{table} 
\caption{\label{benchtable} Computational parameters for the
benchmark simulations in two dimensions.}

\begin{center}
\begin{tabular}{l|c|c}
Quantity & Symbol & Value\\
\hline
Interface thickness & $W_0$ & $1$ \\
Anisotropy & $\epsilon_4$ & $0.025666$, $0.000666$ \\
Effective Anisotropy & $\epsilon_4^e$ & $0.025$, $0.0$ \\
Relaxation time & $\tau_0$ & $1$ \\
Kinetic anisotropy & $\delta_4$ & $0$ \\
Grid spacing & $\Delta x$ & $0.4$ \\
Timestep & $\Delta t$ & $0.003$ \\
Diffusion coefficient & $D$ & $10$ \\
Coupling constant & $\lambda$ & $15.957$ \\
Capillary length & $d_0$ & $0.0554$ \\
Kinetic coefficient & $\beta_0$ & $0$ \\
Undercooling & $\Delta$ & $0.3$ \\
Coarse cell size & $n$ & $4$ \\
Number of walkers per coarse cell & $M$ & $50$ \\
\end{tabular}
\end{center}
\end{table}
\begin{figure}
\centerline{
\psfig{file=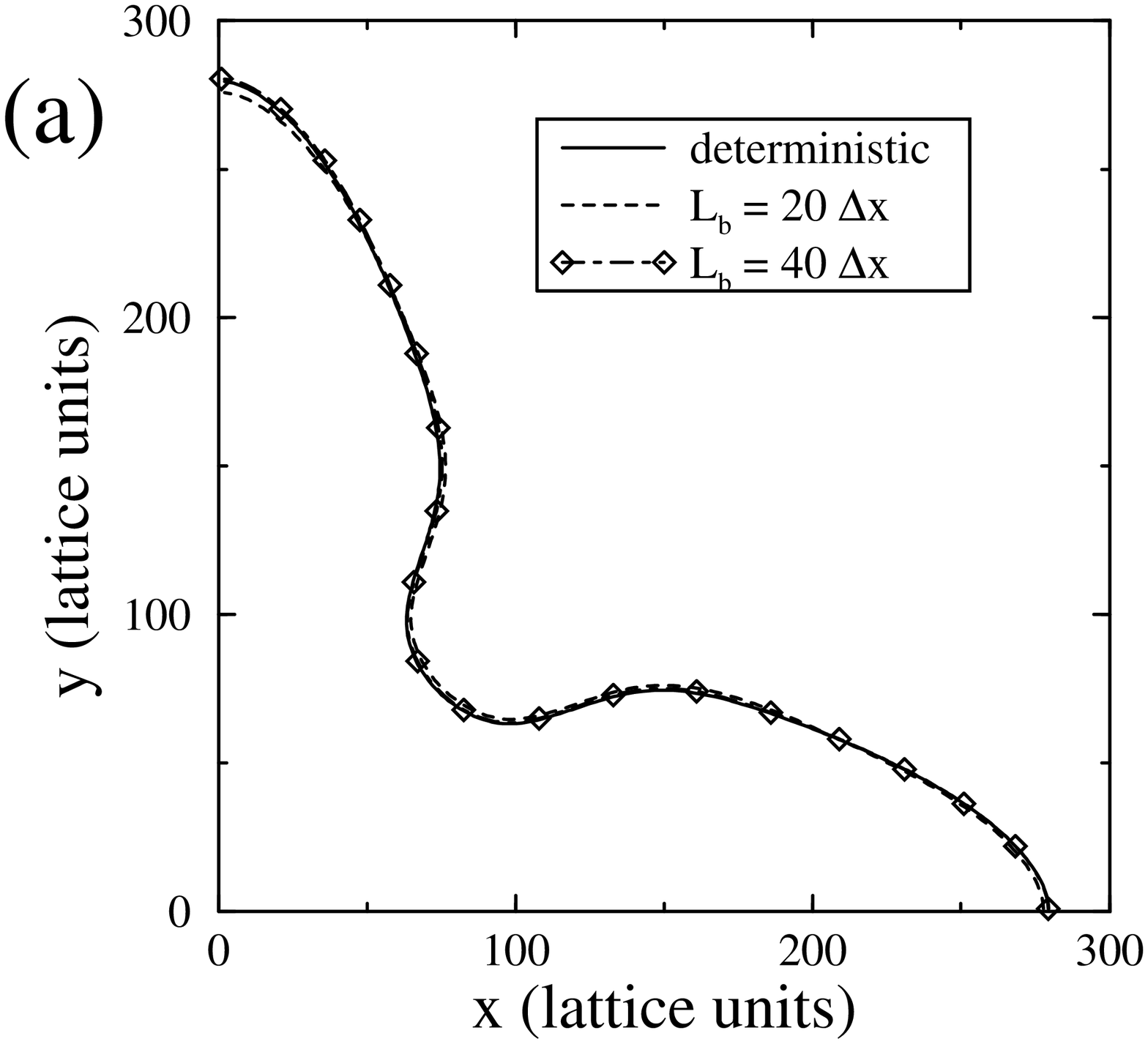,width=.35\textwidth}}
\smallskip
\centerline{
\psfig{file=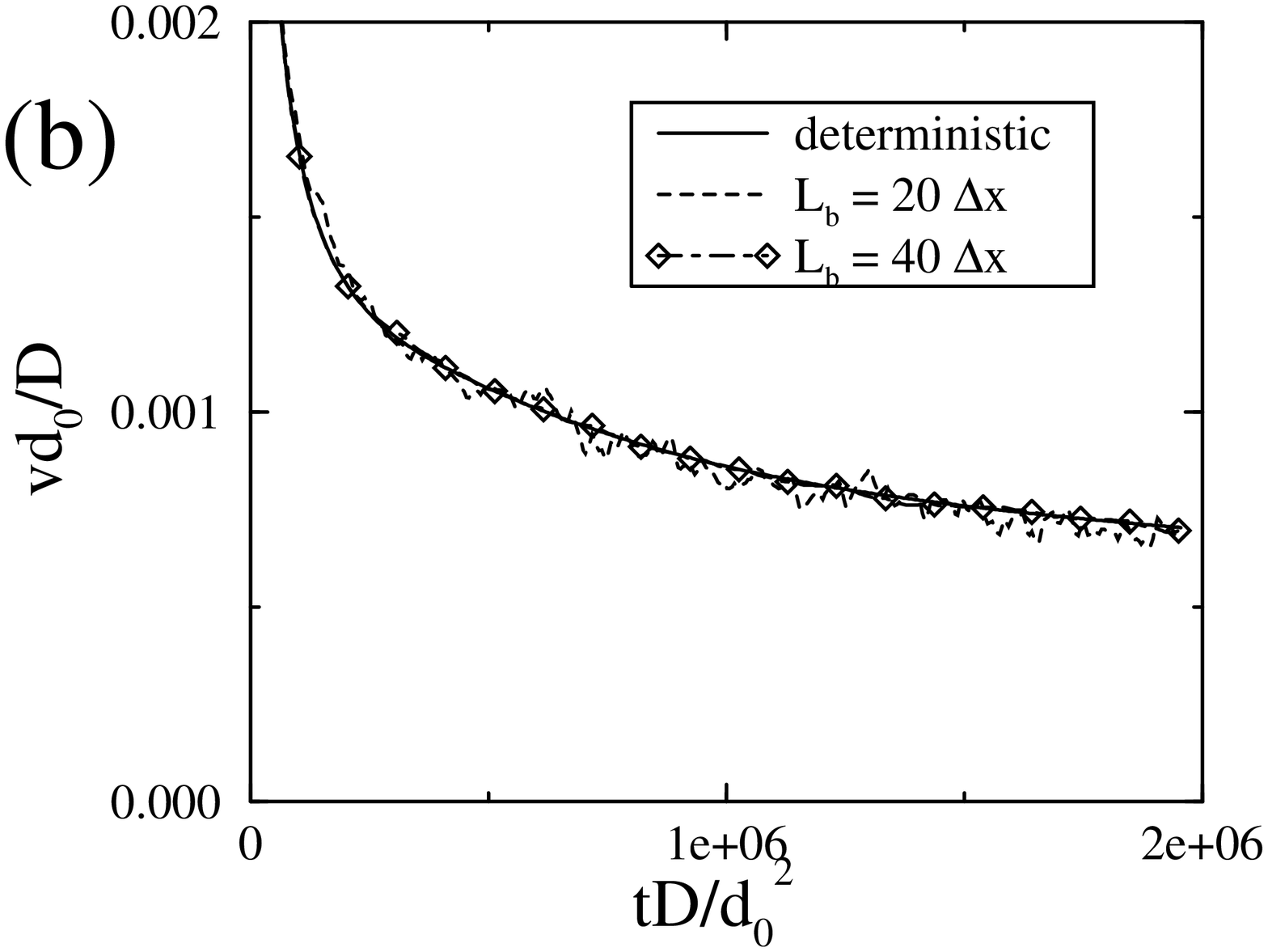,width=.35\textwidth}}
\caption{Comparison of standard (deterministic) phase-field
and random walker method in two dimensions for $\Delta = 0.3$ 
and $\epsilon_4=0.025$. (a) Dendrite shapes, represented
by the contour line $\phi=0$, after 200000
iterations, (b) tip velocity versus time.}
\label{figtest2d}
\end{figure}
We conducted two-dimensional simulations at an intermediate
undercooling, $\Delta=0.3$. At this value of $\Delta$, the
standard phase-field method can still be used to simulate
non-trivial length and time scales of dendritic evolution,
but the length scale of the diffusion field is large enough
to provide a serious test for the random walker method,
i.e. the diffusion length is much larger than the thickness
of the buffer layer. Table \ref{benchtable} shows the 
computational parameters that were used for these tests.
Only the first quadrant was simulated,
with reflecting boundary conditions at $x=0$ and $y=0$.
Fig. \ref{figtest2d}(a) shows a comparison of dendrite
shapes obtained from the standard phase-field and from
our algorithm with different buffer sizes. While the
shapes slightly differ for $L_b/\Delta x = 20$, the
curve for $L_b/\Delta x = 40$ is almost undistinguishable  
from the deterministic shape. Fig. \ref{figtest2d}(b)
shows the velocity of the dendrite tip along the 
$x$-direction, measured over periods of 500 iterations,
versus time. The fluctuations around the deterministic
value are much larger for $L_b/\Delta x = 20$ than for
$L_b/\Delta x = 40$, and for $L_b/\Delta x = 80$ (not
shown) the curve obtained from the stochastic method
is very close to the deterministic data. For comparison,
the diffusion length $2D/v$ at the end of the run is
about $400\,\Delta x$.

\begin{figure}
\centerline{
\psfig{file=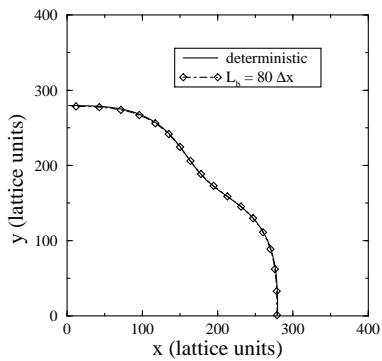,width=.3\textwidth}}
\caption{Comparison of ``dendrite'' shapes without fourfold
anisotropy after 500000 iterations.}
\label{figcirc}
\end{figure}
A particularly sensitive test for the anisotropy of the
conversion process is the growth of a circular germ 
without anisotropy, because such a germ
is unstable against even smallest perturbations. This
can be clearly seen from Fig. \ref{figcirc}: even
though we completely screen the fourfold anisotropy
created by the lattice ($\epsilon_4^e = 0$), the weak
next harmonic of the lattice anisotropy, proportional
to $\cos 8\theta$, destabilizes the circle and leads
to the formation of bulges in the $(21)$- and 
$(12)$-directions. For $L_b/\Delta x = 80$, the stochastic
algorithm perfectly reproduces this trend, and we can
hence conclude that the anisotropy created by the
coarse structure of the conversion boundary is 
negligibly small. Note that the diffusion field
extends to a distance of more than 1000 lattice units at 
the end of this run, which means that the larger 
part of the simulation domain is integrated by the
stochastic method.

\begin{figure}
\centerline{
\psfig{file=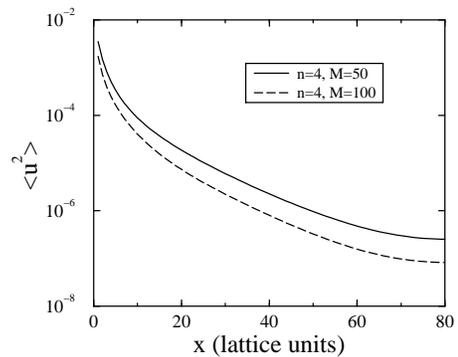,width=.35\textwidth}}
\caption{Variance of temperature fluctuations, 
$\left<u^2\right>$, as a function of the distance 
from the conversion boundary for two values of
the walker parameter $M$.}
\label{figfluct}
\end{figure}
To quantify the numerical noise, we performed 2-d
simulations of the simple diffusion equation in a 
system of $N\times N$ lattice sites with $N=160$. 
One half of the system ($x<0$) was integrated by the
stochastic algorithm, whereas in the other half ($x>0$)
we used a standard Euler algorithm. The conversion
boundary $\Gamma'$ hence coincides with the $y$-axis,
and there is a single column of conversion cells
along this axis. We used $\Delta x=1$,
$\Delta t=0.02$, $D=1$, $\Delta=1$, and applied no-flux 
boundary conditions at $x=\pm N/2$ and periodic boundary
conditions along $y$. The system was initialized 
at $u=0$ everywhere, i.e. in the walker region we 
randomly placed $M$ walkers in each coarse cell. 
When the walkers evolve, fluctuations are created
in the deterministic region, which plays the role of
the buffer layer. We recorded $u^2$ as a function
of $x$ and averaged over a time which is long compared
to the diffusive relaxation time of the system, $N^2/D$.
The results for two different choices of the walker
parameter $M$ are shown in Fig. \ref{figfluct}. 
In an infinite homogeneous system filled 
with walkers, the distribution of the number
of walkers in a given coarse cell is Poissonian,
which means that the fluctuations in the walker
numbers are of order $\sqrt M$. If this scaling
remains valid for the conversion cells in our
hybrid system, we expect $\left<u^2\right> \sim 1/M$ 
close to the conversion boundary, which is indeed 
well satisfied. As shown in Fig \ref{figfluct},
the variance of the temperature fluctuations
rapidly decreases with the distance from the
conversion layer -- by four orders of magnitude
over the distance of $80$ lattice sites. No
simple functional dependence of $\left<u^2\right>$
on $x$ is observed. We expect high spatial and
temporal frequencies to be rapidly damped. A theoretical
calculation of $\left<u(x)^2\right>$ seems possible
but non-trivial because the random variables which
are the sources of the fluctuations in the deterministic
region are correlated in space and time by the exchange
of walkers through the stochastic region and the
diffusion of heat through the deterministic region.
For our present purpose, we can draw two important
conclusions. Firstly, for a reasonable thickness of
the buffer layer, fluctuations are damped by several
orders of magnitude. The residual fluctuations are
much smaller than the thermal fluctuations represented
by Langevin forces that have to be introduced in
the equations of motion to observe a noticeable
sidebranching activity \cite{Karma99}. Indeed,
for sufficiently large buffer layers we always
observe needle crystals without sidebranches. 
Secondly, the fluctuations at the solid-liquid
interface can be reduced both by increasing the
number of walkers and by increasing the thickness
of the buffer layer, which allows to accurately
simulate dendritic evolution with a reasonable
number of walkers.

\begin{table} 
\caption{\label{timetab} Execution times 
of the benchmark simulations for various sets of
computational parameters.}

\begin{center}
\begin{tabular}{l|c|c|c}
 & $M$ & $L_b/\Delta x$ & CPU time (min) \\
\hline
Deterministic & -- & -- & $1950$ \\
  & $50$ & $20$ & $89$ \\
  & $50$ & $40$ & $110$ \\
  & $100$ & $40$ & $119$ \\
\end{tabular}
\end{center}
\end{table}
In Table \ref{timetab}, we compare the run
times of our code on a DEC Alpha 533 MHz workstation along
with the run time of the deterministic phase
field reference simulation. The gain in computational
efficiency is obvious. Increasing the buffer layer
from $L_b=20\,\Delta x$ to $L_b=40\,\Delta x$
reduces the amplitude of the temperature fluctuations
at the solid-liquid interface by more than an order of
magnitude, whereas the computation time increases by
only $25$\%. Comparing the runs with different
values of $M$, we see that the walker part of the
program accounts only for a small part of the total
run time.

From these results, we can conclude that the computational
effort that has to be invested to simulate a given time
increment scales approximately as the size of the fine
grid region, i.e. as the size of the dendrite. This is a
major advance with respect to the standard phase-field implementation
on a uniform grid,
where the computation time scales with the volume enclosing the
diffusion field. The spatial and temporal scales of
dendritic evolution that can be simulated with our
method are hence limited by the integration of the 
phase-field equations on the scale of the dendrite.

\begin{figure}
\centerline{
\psfig{file=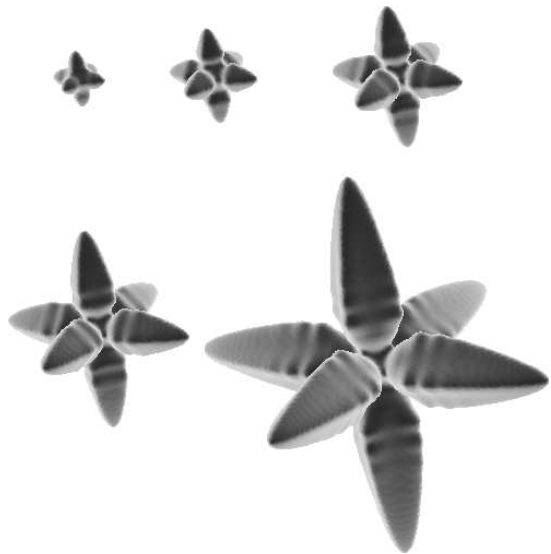,width=.4\textwidth}}
\smallskip
\caption{Snapshots of a three-dimensional dendrite at
$\Delta=0.1$ after 60000, 120000, 200000, 300000,
and 650000 timesteps (from top left).}
\label{figsnap}
\end{figure}

\begin{figure}
\centerline{
\psfig{file=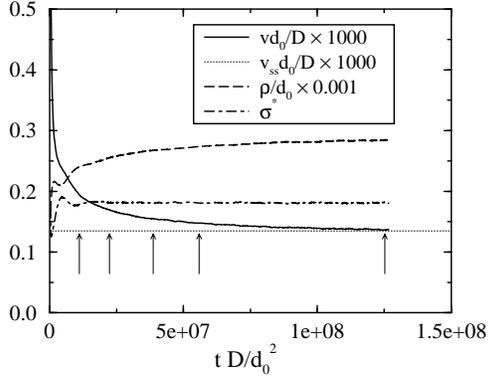,width=.4\textwidth}}
\caption{Tip velocity, tip radius and selection parameter
versus time for the run of Fig. \protect\ref{figsnap}.
Arrows mark the times of the snapshot pictures.
Length and time are rescaled by $d_0$ and $d_0^2/D$,
respectively. The steady-state velocity $v_ss$ was
calculated using a boundary integral 
method \protect\cite{Lee99}.}
\label{figplot}
\end{figure}

All the data shown so far are for two-dimensional
simulations. We repeated similar tests in three 
dimensions and obtained comparable results for the 
quality of the solution and the efficiency of the 
code. We will not display the details of these 
comparisons here, but rather show an example of
a three-dimensional simulation under realistic
conditions to demonstrate that our method is capable 
of yielding quantitative results in a regime of
parameters that was inaccessible up to now.
In Fig. \ref{figsnap}, we show snapshot pictures
of a three-dimensional dendrite growing at an
undercooling of $\Delta=0.1$ and for a surface
tension anisotropy $\epsilon_4^e=0.025$, which is the
value measured for PVA \cite{Muschol92}. The other
computational parameters are $W_0=1$, $\Delta x=0.8$, 
$\epsilon_4=0.0284$, $\tau_0=0.965$, $\delta=0.0364$, 
$n=4$, $M=50$, $L_b/\Delta x = 48$, $D=24$, $\Delta t=0.004$, 
and $\lambda=39.6$ (giving $d_0 = 0.0223$, $\beta_0=0$),
and the simulation was started from a homogeneously
undercooled melt with an initial solid germ of radius
$r=2\Delta x$ centered at the origin.
During the run, we recorded the velocity $v(t)$ and
the radius of curvature $\rho(t)$ of the dendrite
tip. The latter was calculated using the method 
described in Ref. \cite{Karma98}. With these two
quantities, we can calculate the time-dependent
tip selection parameter
\be
\sigma^*(t) = {2Dd_0\over \left[\rho(t)\right]^2 v(t)}.
\ee
The results are shown in Fig. \ref{figplot}.
In the initial stage during which the arms emerge
from the initial sphere, growth is very rapid. 
Subsequently, the tips slow down while the 
diffusion field builds up around the crystal.
At the end of the run, the velocity has almost converged
to a constant value that is in excellent quantitative
agreement with the velocity predicted by the boundary integral
solution of the sharp-interface steady-state growth equations 
assuming an axisymmetric surface energy and tip shape (i.e. the 
most accurate numerical implementation of solvability theory 
to date \cite{Lee99}). This velocity is also 
in reasonably good agreement with the velocity predicted 
by the linearized solvability theory of Barbieri and Langer
\cite{Barbieri89}, even though the actual tip radius in both 
the phase-field simulation and the boundary integral calculation 
differ from the tip radius of the paraboloidal shape assumed 
in this theory. A more detailed discussion of 
this point and the entire steady-state tip morphology 
can be found in Ref. \cite{Tip99}.

Remarkably, the selection constant $\sigma^*$ becomes
almost constant long before the velocity and the tip
radius have reached their steady-state values. This
is in good agreement with the concepts of solvability
theory, which stipulates that the selection of the
tip parameters is governed by the balance between
the anisotropic surface tension and the local diffusion
field at the tip. To establish the correct local balance,
diffusion is necessary only over a distance of a few
tip radii, whereas the buildup of the complete
diffusion field around an arm requires heat transport
over the scale of the diffusion length, $D/v$. Our
simulation shows that $\sigma^*$ indeed becomes essentially
constant soon after the formation of the primary arms.
This fact can be used to derive scaling laws for the
evolution of the dendrite arms at low undercooling
during the transient that leads to steady-state
growth \cite{Plapp99}. Finally, even at the end of
the simulations, where the dendrite arms are well
developed, no sidebranches are visible. We repeated
the same simulation for different thickness of the
buffer layer, and observed no changes in the morphology.
Tiny ripples can in fact be seen close to the base of
the dendrite shaft, but the amplitude of these
perturbations does not depend on the noise strength.
We therefore believe that this is rather a deterministic
instability due to the complicated shape of the
dendrite base that a beginning of noise-induced 
sidebranching. In summary, there are at present no
indications that the noise created by the walkers
has a noticeable effect on the morphological evolution.

\section{Conclusions}
\nobreak
\noindent
We have presented a new computational approach for
multi-scale pattern formation in solidification.
The method is efficient, robust, precise, easy to implement in 
both two and three dimensions, and parallelizable. Hence, it 
constitutes a powerful alternative to state of the art
adaptive meshing and finite element techniques. We have
illustrated its usefulness by simulating dendritic growth 
of a pure substance from its undercooled melt
in an infinite geometry. Due to the fact that only
a very limited amount of ``geometry bookkeeping''
is required, our method can be easily adapted to other
experimental settings, such as directional solidification.
In addition, the DMC algorithm is not limited to the
present combination with the phase-field method, but 
can be used in conjunction with any method to solve 
the interface dynamics, as long as the diffusion equation 
is explicitly solved. The adaptation of our method to other
diffusion-limited free boundary problems is straightforward;
problems with several diffusion fields can be handled by
introducing multiple species of walkers.

In view of the results presented here,
there is a realistic prospect for direct simulations
of solidification microstructures for experimentally
relevant control parameters. An especially interesting prospect 
is to combine our method with a recently developed approach to
{\it quantitatively} incorporate thermal fluctuations \cite{Karma99}
in the phase-field model. Such an extension should make it possible
to test noise-induced sidebranching theories \cite{Langer87,Brener95} 
in three dimensions and for an undercooling range where detailed 
measurements of sidebranching characteristics are available 
\cite{Huang81,Bisang95,Dougherty87,Li98}. 
If thermal noise in the liquid region outside the
buffer layer turns out to be unimportant for sidebranching, 
the straightforward addition of Langevin forces as in 
Ref. \cite{Karma99} in the finite-difference region 
(i.e. the buffer region plus the solid) should suffice for
this extension. In contrast, if the noise from this region is
important, a method to produce the correct level of noise in 
the walker region will need to be developed. Work concerning 
this issue is currently in progress. 

To conclude, let us comment on some possible extensions and
improvements of our method which will be necessary to address 
certain questions. Firstly, we have described the method
here using an explicit integration scheme on the fine grid in 
the inner region, which enforces rather small time steps. We 
also tested an alternating direction Crank-Nicholson scheme
in 2-d, which speeds up the calculations but makes it
necessary to introduce corrective terms at the conversion
boundary to guarantee the local heat conservation. 
Secondly, for the moment we use the stochastic algorithm 
only at the {\it exterior} of the dendrite; for other
geometries, such as directional solidification where
the volumes of solid and liquid are comparable, it might 
be useful to introduce a second stochastic region in the 
solid. It would also be desirable to combine our algorithm 
with more efficient memory managing techniques to overcome
the limitations due to storage space. Finally, a
completely open question is whether it is possible to
combine our stochastic algorithm with a suitable method
for simulating hydrodynamic equations. This would open
the way for studies of the influence of convection on dendritic
evolution at low undercooling, thereby extending in a non-trivial
way recent studies that have been restricted 
to a relatively high undercooling regime \cite{Crisandme}.

\acknowledgments

This research was supported by U.S. DOE Grant 
No. DE-FG02-92ER45471 and benefited from 
computer time at the National Energy Research 
Scientific Computing Center (NERSC) at
Lawrence Berkeley National Laboratory and
the Northeastern University Advanced Scientific
Computation Center (NU-ASCC). We thank Youngyih
Lee for providing the boundary integral results
and Flavio Fenton for his help with the 
visualization. Fig. \ref{figsnap} was created
using Advanced Visual Systems' AVS.

\appendix

\section{Phase-field method}
\nobreak
\noindent
We will briefly outline the main features of the
phase-field method used for our simulations. More
details can be found in Ref. \cite{Karma98}.

The starting point is the free energy functional,
Eq. (\ref{freeen}), together with the equations
of motion for the phase field and the temperature
field, Eqs. (\ref{pfpsi}) and (\ref{pfu}). The
free energy density in Eq. (\ref{freeen}) is
chosen to be of the form
\be
f(\psi,u)=-{\psi^2\over 2}+{\psi^4\over 4}+\lambda\,u\,\psi\,
\left(1-2{\psi^2\over 3}+{\psi^4\over 5}\right).
\label{freeden}
\ee
This function has the shape of a double well, with minima 
at $\psi = \pm 1$ corresponding to the solid and the liquid
phases, respectively. Here, $u$ is the dimensionless 
temperature field, $\lambda$ is a dimensionless coupling
constant, and the term proportional to $u$ on the RHS of 
Eq. (\ref{freeden}) ``tilts'' the double well in order to favor 
the solid (liquid) minimum when the temperature is below (above)
the melting temperature. The coefficient $W(\hat n)$ of
the gradient term in the free energy (\ref{freeen}) determines 
the thickness of the diffuse interface, i.e. the scale on
which the phase field varies rapidly to connect the
two equilibrium values. In addition, $W$ is related to
the surface tension, and exploiting its dependence on the
orientation of the interface allows to recover the 
anisotropic surface tension of Eq. (\ref{ani}) by choosing
\be
W(\hat n) = W_0 {\gamma(\hat n)\over \gamma_0}.
\label{wthet}
\ee
The orientation $\hat n$ is defined in terms of the phase field by
\be
\hat n = {\vnab \psi\over | \vnab\psi|}\,.
\ee
Note that this dependence of $W$ on $\psi$ has to be taken
into account in performing the functional derivative, such
that the explicit form of Eq. (\ref{pfpsi}) becomes
\begin{eqnarray}
\tau(\hat n)\partial_t\psi & = & 
   [\psi - \lambda u(1-\psi^2)](1-\psi^2) + 
   \vnab\cdot[W(\hat n)^2\vnab\psi] \nonumber \\
  & & \mbox{} + \partial_x \left({|\vnab\psi|}^2 W(\hat n)
       {\partial W(\hat n) \over \partial(\partial_x \psi)}\right) \nonumber \\
  & & \mbox{} + \partial_y \left({|\vnab\psi|}^2 W(\hat n)
       {\partial W(\hat n) \over \partial(\partial_y \psi)}\right) \nonumber \\
  & & \mbox{} + \partial_z \left({|\vnab\psi|}^2 W(\hat n)
       {\partial W(\hat n) \over \partial(\partial_z \psi)}\right).
\label{pfnum}
\end{eqnarray}

Next, we need to specify the orientation-dependent 
relaxation time $\tau(\hat n)$ of the phase-field.
In analogy with Eqs. (\ref{wthet}) and (\ref{ani})
we choose
\be
\tau(\hat n) = \tau_0\,(1-3\delta_4)\,
\left[1+\frac{4\delta_4}
{1-3\delta_4}\left(n_x^4+n_y^4+n_z^4\right)\right],
\label{anikin}
\ee
where $\delta_4$ is the kinetic anisotropy.

The phase-field equations can be related to the original
free boundary problem by the technique of matched asymptotic
expansions. Details on this procedure can be found in
Ref. \cite{Karma98}. As a result, we obtain expressions
for the capillary length and the kinetic coefficient
in terms of the phase-field parameters $W_0$ and 
$\tau(\hat n)$:
\be
d_0 = {a_1W_0\over \lambda}
\label{d0}
\ee
\be
\beta(\hat n) = {a_1\over\lambda}{\tau(\hat n)\over W_0}
   \left(1-a_2 \lambda{W(\hat n)^2\over D\tau(\hat n)}\right),
\label{beta}
\ee
where $a_1 = 0.8839$ and $a_2 = 0.6267$ are numerical 
constants fixed by a solvability condition.
There is an important difference between this result and
earlier matched asymptotic expansions of the phase-field
equations, due to a different choice of the expansion
parameter. If the coupling constant $\lambda$ is used
as the expansion parameter, the first order
in $\lambda$ gives only the first term in 
Eq. (\ref{beta}), while the complete expression
is the result of an expansion to first order
in the interface P{\'e}clet number, which is defined
as the ratio of the interface thickness and a relevant
macroscopic scale of the pattern (local radius of curvature
or diffusion length). An important consequence of
Eq. (\ref{beta}) is that the kinetic coefficient and
its anisotropy can be set to arbitrary values by a 
suitable choice of $\lambda$ and $\tau(\hat n)$,
and in particular we can achieve vanishing kinetics
($\beta(\hat n) = 0$). Note that for a $\tau(\hat n)$
as given by Eq. (\ref{anikin}), the kinetic coefficient
cannot be made to vanish simultaneously in all directions,
but for small anisotropies choosing $\delta_4=2\epsilon_4$
is a sufficiently accurate approximation.
Furthermore, the ratio $d_0/W_0$
can be decreased without changing the kinetics by 
simultaneously increasing $\lambda$ and the diffusivity $D$. 
This method dramatically increases the computational
efficiency of the phase-field approach, because the
interface width $W_0$ determines the grid spacing which
must be used for an accurate numerical solution. For
a physical system with fixed capillary length $d_0$,
the number of floating point operations necessary to
simulate dendritic evolution for some fixed time interval
and system size scales $\sim(d_0/W_0)^{d+3}$ for the choice
of phase-field parameters where the interface kinetics vanish
(i.e. $D\tau/W_0^2 \sim \lambda \sim W_0/d_0$), where $d$ is the 
spatial dimension \cite{Karma98}.

We integrate the phase-field equations on a cubic grid 
with spacing $\Delta x$, All spatial derivatives
are discretized using $(\Delta x)^2$-accurate finite 
difference formulas, and timestepping is performed by a 
standard Euler algorithm. The use of a regular grid
induces small anisotropies in the surface tension and
the kinetic coefficient. These effects have been precisely
quantified in Ref. \cite{Karma98}. Since the grid has the same 
symmetry as the crystal we want to simulate, the presence of
the lattice simply leads to small shifts in the surface
tension anisotropy and in the kinetic parameters. 
For example, we obtain an effective 
surface tension anisotropy $\epsilon_4^e$ which is slightly 
smaller than the ``bare'' value $\epsilon_4$. Evidently,
the use of this method restricts the simulation to crystals
with symmetry axes aligned to the lattice, but this is
not a severe limitation in the present study which focuses 
on the growth of single crystals.

\end{document}